# Planning Fallacy or Hiding Hand: Which Is the Better Explanation?

By Bent Flyvbjerg[*]

*Abstract*: This paper asks and answers the question of whether Kahneman's planning fallacy or Hirschman's Hiding Hand best explain performance in capital investment projects. I agree with my critics that the Hiding Hand exists, i.e., sometimes benefit overruns outweigh cost overruns in project planning and delivery. Specifically, I show this happens in one fifth of projects, based on the best and largest dataset that exists. But that was not the main question I set out to answer. My main question was whether the Hiding Hand is "typical," as claimed by Hirschman. I show this is not the case, with 80 percent of projects not displaying Hiding Hand behavior. Finally, I agree it would be important to better understand the circumstances where the Hiding Hand actually works. However, if you want to understand how projects "typically" work, as Hirschman said he did, then the theories of the planning fallacy, optimism bias, and strategic misrepresentation – according to which cost overruns and benefit shortfalls are the norm – will serve you significantly better than the principle of the Hiding Hand. The latter will lead you astray, because it is a special case instead of a typical one.

[*] Bent Flyvbjerg is the BT Professor and Chair of Major Programme Management at Oxford University's Saïd Business School.

I want to thank the editor for giving me the opportunity to respond to the comments made above by Philipp Lepenies, Graham Room, and Lavagnon Ika on my paper "The Fallacy of Beneficial Ignorance: A Test of Hirschman's Hiding Hand," printed in *World Development*, vol. 84. First, I will briefly mention some of the points on which I agree with my critics. Second, I will respond to their main criticisms.

I agree regarding the richness and originality of Hirschman's work and that he was a leading 20th century intellectual and economist, well worth reading today. His ideas may still open up a range of productive and unsuspected new vistas, as maintained by Room (pp. 1, 2). I also agree that many of Hirschman's observations regarding the principle of the Hiding Hand, and the fundamentally anti-rationalist worldview they represent, are highly innovative, for instance that success may be more often stumbled upon than carefully planned for, as argued by Lepenies (p. 5). Or that the World Bank's push for cost-benefit analysis as a synoptic project evaluation method might have some use, but was ultimately a misguided attempt at comprehensive quantification, as pointed out by Ika (p. 38). Finally, I agree that "the Hiding Hand is a possible empirical occurrence and it does happen," as observed by Ika (p. 4), i.e., creative project managers are sometimes able to generate benefit overruns that are larger than cost overruns, securing the viability of projects.

My critics have the grace to similarly point out many instances where they agree with what I say. For example, Ika (p. 4) acknowledges that until he began working on his rejoinder to my paper he was under the impression that the principle of the Hiding Hand was not empirically testable, but now he shares my view that the principle is not only a theory for explanation but also an empirically testable hypothesis. Room (p. 2) says he finds it difficult to disagree with me that Hirschman had insufficient empirical evidence for affirming the Hiding Hand, just as Ika (p. 14) agrees with the limitations I identify for Hirschman's work, including his small sample of 11 projects. "Admittedly, these methodological limitations plagued Hirschman's findings," Ika (p. 14) concedes. Finally, Ika (p. 4) further agrees with my critique of Hirschman for having sampled on the independent variable in his study of the Hiding Hand.

When it comes to dissent, my critics disagree not only with me, but also among themselves, which weakens their critique, needless to say. For instance, Ika (p. 4) accepts my claim that the principle of the Hiding Hand is a testable theory, as mentioned above, whereas Lepenies (p. 7) rejects this. Furthermore, Lepenies (p. 2) takes at face value Hirschman's success story about the paper mill in Pakistan, and uses this to argue that the Hiding Hand works. In contrast, Ika (p. 8, note 6) rightly observes that the paper mill turned out to be "a disastrous development failure." This is the case, too, for several other projects that Hirschman identified as successes and took to support the Hiding Hand, but that turned into failures shortly after Hirschman completed his study, falsifying the Hiding Hand. Surely this is a problem, as pointed out in my paper. But Lepenies glosses it over, as did Hirschman when confronted with the facts.

**What is Theory?**

Lepenies (p. 6) further confuses normative and explanatory theory when he quotes Hirschman as having a "dislike for general principles" and theory. Hirschman, as quoted by Lepenies, was explicitly talking about normative theory, in terms of "prescriptions," "recipe[s]," and "therapy" (p. 6; Hirschman 1998: 88, 110). It is correct that Hirschman was against theory used in this manner, i.e., for design, where law-like ideas are used to prescribe and plan social action, or even the social order. Hirschman generally saw such social engineering – whether driven by experts or revolutionaries – as highly problematic and bound to fail. But Hirschman had no issue with explanatory theory, not for himself and not for other social scientists. And the principle of the Hiding Hand was developed by Hirschman as explanatory theory, with truth claims and causal mechanisms. By not distinguishing between normative and explanatory theory, as Hirschman does, Lepenies misses this important point and is led to wrongly claim that Hirschman was against theory as such, when he was only against normative theory. Hirschman (1994: 277–78) explicitly referred to his work as "theory building" and added, "I bristle a bit when I am pigeonholed as 'atheoretical' or 'antitheoretical'." Hirschman would bristle, and feel pigeonholed, if he read Lepenies's take on his work.

Room (p. 2) claims I have "a particular view of theory – as hypotheses that can be subjected to quantitative empirical assessment in large datasets." This is incorrect. If Room had read my paper more carefully, he would not have had to guess at my view of theory, and get it wrong. He would have found that I explicitly write in my paper (p. 186, n. 10), "[t]he term 'theory' is here used to denote an idea, or a system of ideas, used to account for or explain a situation." This definition is not limited to "quantitative empirical assessment" or "large datasets." It would apply just as well to qualitative phenomena and qualitative validity assessments. My definition of theory is therefore wider than the tired either/or of quantitative versus qualitative assessment that Room seems to allude to in his false critique. It is correct that I use quantitative data to test the principle of the Hiding Hand. But this is because the Hiding Hand makes truth claims in terms that are best tested quantitatively, for instance the claim that "people typically take on and plunge into new tasks because of the erroneously presumed absence of a challenge," in Hirschman's (1967, p. 13) words. By using the term "typically" this claim explicitly indicates that the depicted behavior applies more often than not, which is a specific truth claim that demands quantitative testing for its verification or rejection, as done in my paper. Or when Hirschman (1967, p. 15) states that according to the Hiding Hand "costs are underestimated and investment decisions activated in consequence," which again is a truth claim that lends itself to empirical, quantitative test (the costs are underestimated or not, and the investment decisions are activated or not), as done in my paper. This is not me imposing a quantitative framework on Hirschman and the principle of the Hiding Hand, as Room suggests. This is me testing the principle on Hirschman's own terms, chosen by him in his formulation of his truth claims, like those quoted above.

**Will Wider Impacts Save the Day?**

Room (p. 4) appeals to "dynamic linkages" and a "wider development calculus" in an attempt to problematize the focus on direct benefits and costs in my tests of the Hiding Hand. Ika (p. 5) similarly conjures up "unintended effects of projects" and "full life-cycle costs and benefits" as "critical for a valid assessment of the Hiding Hand." Such wider impacts are often invoked in attempts to justify projects that may not be viable in terms of direct benefits and costs. It would be nice – and good academic practice – if proponents of the wider-impacts argument, including Room and Ika, would provide empirical evidence that wider impacts are in fact significant and may move the needle from non-viable to viable, if included in project appraisal. The fact is that proponents rarely provide such evidence, and for good reason. The evidence does not exist. Roger Vickerman – a leading expert on wider impacts – recently did a study of the state-of-the-art of research in this area. Choosing transportation infrastructure projects as his example, as such projects are often argued to have large wider impacts, he looked at the existing evidence and concluded (Vickerman 2017, pp. 401-402):

1. Positive wider impacts, where they exist, typically account for an additional 10-20 percent of benefits.[1]
2. Positive wider impacts are not guaranteed for every project.
3. Where positive wider impacts do exist for some geographical regions they could be negative for others, reducing the aggregate effect.
4. The common assumption is deeply problematic that wider benefits will come to the rescue of a project which is marginal on the basis of its direct benefits and costs.
5. Only in "very particular cases" are wider benefits likely to rescue a project from non-viability.
6. Wider impacts were never intended to be a cure for investment appraisals, especially marginal ones, but only a way to ensure completeness.
7. Some wider impacts, as currently measured, are argued by some to be a "mirage essentially involving double counting of direct benefits," in the words of Vickerman.

Room, Ika, and other proponents of the wider-impacts argument may hope that such impacts will come to the rescue of non-viable projects, and of the Hiding Hand. But hope is not a useful strategy in scholarship. Evidence is. And given the available evidence on wider impacts there is no indication that their inclusion would significantly alter my conclusions about Hirschman's Hiding Hand being atypical; quite the opposite.

**Strawman, What Strawman?**

Ika (p. 15) states that the Hiding Hand "fits well with 'optimism bias'." This must be an oversight on Ika's part. The Hiding Hand says that both costs and benefits will be underestimated in project appraisal, whereas optimism bias predicts that costs will be underestimated while benefits will be overestimated. Empirical tests show that the theory of optimism bias, as formulated by Daniel Kahneman and others, is sound: its predictions are accurate in a majority of cases. Conversely, empirical tests show that the principle of the Hiding Hand is unsound: its predictions are inaccurate in a majority of cases, as documented in my paper. The Hiding Hand and optimism bias therefore do not fit each other well, as Ika says they do, but are directly opposed regarding the prediction of benefits.

Ika (pp. 28-29) further claims that I ignore Hirschman's considerations on "project difficulties" and "problem-solving abilities" in my depiction of the Hiding Hand, and that therefore my version of the Hiding Hand is a "weak version," compared to that of Hirschman. Ika (p. 29) goes so far as to call my alleged weaker version of the Hiding Hand a "straw man." Again this is wrong. I do not ignore project difficulties and problem-solving abilities and I agree with Hirschman and Ika that they are important for understanding project management. To the extent that difficulties have impacted the projects in my dataset this is taken into account in my measurement of the projects' costs and benefits, unless we are talking about the wider impacts dealt with above. Difficulties impact projects in two ways, either as (a) an increase in costs, for instance where a project proved more difficult to build than anticipated, or (b) a reduction in benefits, for example where a project was delivered late or proved more difficult to operate than expected. Not only are such difficulties included in my analysis in the measurement of costs and benefits, they are in accordance with Hirschman's view of the relationship between project difficulties on one hand and project costs and benefits on the other. Similarly for problem-solving abilities, which may impact projects by (a) lowering costs, for instance where a cheaper delivery method than expected was found, or by (b) increasing benefits, for example where ways were found to engage more users than anticipated. Again, such problem-solving abilities are included in my analysis in the measurement of costs and benefits, and again they are in accordance with Hirschman's view of the relationship between problem-solving abilities on one hand and project costs and benefits on the other. The only straw man here is Ika's portrayal of my depiction of the Hiding Hand.

**Data and "Data"**

Finally, Ika (pp. 30-36) presents a remarkable set of data together with conclusions that would be truly revolutionary, in the Kuhnian sense, were they right. Ika's data are from a survey in which World Bank project supervisors were asked to recall a completed or nearly-completed project that they know and then assess how they perceive different aspects of the project's performance on a seven-point scale. On the basis of recording such perceived performance for 161 projects, Ika (p. 32) concludes that the planning fallacy writ large (a.k.a. the Malevolent Hiding Hand)[2] is present in only 3 percent of his projects, in contrast to 80 percent for the 2,062 projects in my study. Ika further concludes that Hirschman's Hiding Hand is four times as prevalent in his data as the planning fallacy. This contrasts with the exact opposite result for my data, where the planning fallacy is four times more prevalent than the Hiding Hand. That is a very large difference, in both size and direction, and both results cannot be true for the wider project population, needless to say. Ika takes his result to disprove my analysis. "Flyvbjerg's claim that the Hiding Hand is typically *less* empirically common than the Planning Fallacy Writ Large turns out to be wrong," concludes Ika (p. 32, emphasis in original).

Let us pause, however, and contemplate the implications of Ika's findings. If he is right, what he has done here is falsify in one fell swoop (a) Nobel Prize-winning theories in behavioral science, including the planning fallacy and optimism bias, argued and proven by Daniel Kahneman, Amos Tversky, and many others (Kahneman and Tversky 1979, Gilovich et al. 2002), and (b) a large body of empirical research in management and planning, which documents that project cost overruns and benefit underruns are significantly more common than cost underruns and benefit overruns (Pickrell 1990, Flyvbjerg et al. 2002, Altshuler and Luberoff 2003, Dantata et al. 2006, Lee 2008, Siemiatycki 2009, Ansar et al. 2014, Flyvbjerg 2016). This would be a revolutionary finding, indeed, as it would topple several well-established research paradigms across the disciplines of economics, management, and planning. If this sounds too good to be true, it is because it is. Ika is wrong for the following reasons.

First, Ika is committing the age-old error of comparing apples and oranges. His study is based on recalled, perceived project performance as subjectively reported by project supervisors on a simple seven-point scale. My study (and most academic studies in this field) is based on actual project performance, measured by the difference between estimated and actual costs and benefits, measured consistently across projects following international standards (Flyvbjerg et al. 2002, Flyvbjerg 2005). To compare recalled, subjectively perceived performance with actual performance, as Ika does, is likely to entail error. For instance, behavioral science has shown that people, including experts, generally perceive and remember outcomes as more positive than they actually are or were (Gilovich et al. 2002). This is part of optimism bias and is likely to have influenced the results of Ika's analysis. It would certainly explain the striking difference between Ika's results and the rest of the field, including my study. Furthermore, theories of strategic misrepresentation say that project proponents have an interest in misrepresenting their projects to appear more successful than they actually are, for instance to secure funding or a positive image. This kind of spin is all-too-human and very common. For example, by deftly rebaselining the budget of the London 2012 Olympics the organizers made a large cost overrun appear to be a small underrun and succeeded in making the media and the public believe in their misrepresentation (Flyvbjerg et al. 2016, p. 2). The proponents of the Øresund bridge between Denmark and Sweden similarly made a cost overrun disappear and magically balanced their budget, again by means of rebaselining, which is the oldest and most-used trick in the book (Flyvbjerg et al. 2003). If you repeat this type of false claim often enough, people start believing it. In time you may even believe it yourself; self-deception is common and is likely to have biased Ika's data. On this background, it is surprising, and disconcerting, that he would rely in his analysis on subjective recollections about success and not systematically reflect on the important and well-documented sources of error and bias this entails, and what they mean to his conclusions.

Second, and worse, Ika's methodology seems manipulated – "knowingly or unknowingly" (p. 29) – to verify the Hiding Hand. When purporting to measure success and failure, Ika (pp. 31-32) divides his seven-point scale into two separate sections, with 1-3 (three choices) signifying project failure and 4-7 (four choices) signifying project success. With this division, even if responses were distributed randomly across the seven-point scale, the outcome would be a bias for concluding that projects are successes on average, which – surprise – happens to support the Hiding Hand. This outcome is not a characteristic of the projects Ika studied, but purely a result of the biased methodology he designed for the study. It is difficult to believe Ika would not have been aware of this self-made bias and how it affects his conclusions, but nowhere does he warn his readers against it, which makes his analysis highly dubious on this point.

In sum, due to these errors and biases, Ika's so-called "data" – and the "findings" built on them – are methodological artifacts that say little about reality and much about a relaxed attitude to validity and truth that has no place in scholarship.

**How Projects Typically Work**

Let me end on a more positive note by reiterating that I agree with Lepenies, Room, and Ika that the Hiding Hand exists, i.e., sometimes benefit overruns outweigh cost overruns. Specifically, I show this happens in one fifth of projects. But that was not the main question I set out to answer. My main question was whether the Hiding Hand is "typical," as claimed by Hirschman. I show this is not the case, with 80 percent of projects not displaying Hiding Hand behavior. Finally, I agree with Ika (p. 41) that it would be important to better understand the circumstances where the Hiding Hand actually works, and I believe Hirschman's original thoughts on the Hiding Hand would be a good starting point for developing such an understanding. However, if you want to understand how projects "typically" work, as Hirschman said he did, then the theories of the planning fallacy, optimism bias, and strategic misrepresentation – according to which cost overruns and benefit shortfalls are the norm – will serve you significantly better than the principle of the Hiding Hand. The latter will lead you astray, because it is a special case instead of a typical one, and taking wider impacts into account is unlikely to change this. Nothing of what my critics say above weakens this conclusion.

**Notes**

[1] This does not take into account negative wider impacts, like environmental and social costs, which are often substantial for large infrastructure projects, and which, if included, would reduce the aggregate effect of wider impacts.

[2] The planning fallacy was originally formulated by Kahneman and Tversky (1979) and Buehler et al. (1994) to apply to estimates of completion times and schedules. The concept was later expanded by Lovallo and Kahneman (2003), Flyvbjerg (2008), and Flyvbjerg and Sunstein (2017) to also include the costs and benefits of decisions. Thus the "writ large." In what follows the term "planning fallacy" is used as shorthand for the planning fallacy writ large, i.e., the tendency for people to underestimate the costs and completion times of projects and overestimate their benefits.